\documentclass{preprint}
\usepackage{times}
\usepackage{mhequ}
\usepackage{mhenvs}
\usepackage{Symbols}

\def\prob{{\bf P}}
\def\expect{{\bf E}}

\def\XXX#1{{\hfill\break\null\kern
-2truecm
{\tt ********** #1}
}} 

\newtheorem{assump}[lemma]{Assumption}
\newcommand{\loc}{{\,\mathrm{loc}}}

\begin{document}
\title{Stationary solutions for a model of\\ amorphous thin-film growth}
\author{Dirk Bl\"omker$^1$ and Martin Hairer$^2$}
\institute{
  $^1$ Institut f\"ur Mathematik, RWTH Aachen, Germany\\
  \phantom{$^1$ }Email: {\tt bloemker@instmath.rwth-aachen.de}\\
  $^2$ D\'epartement de Physique Th\'eorique, University of Geneva, Switzerland\\ 
  \phantom{$^2$ }Email: {\tt Martin.Hairer@physics.unige.ch}}
\maketitle

\begin{abstract}
We consider a class of stochastic partial differential equations arising as a model
for amorphous thin film growth. Using a spectral Galerkin method, 
we verify the existence of stationary mild solutions, although the specific nature of the nonlinearity prevents us from showing the uniqueness of the solutions as well as their boundedness (in time).
\end{abstract}

\section{Introduction}

This paper shows the existence of a stationary solution for
a stochastic partial differential equation (SPDE), 
where the solutions may not form
a Markov semigroup due to the lack of uniqueness. We consider the family of equations
\begin{equ}[e:mainequ]
\d_t u = -\d_x^4 u + \nu\d_x^2 u - \d_x^2 (\d_x^{} u)^2 + \xi(x,t)\;,\qquad \nu\in\R\;,
\end{equ}
for a real-valued scalar $u(t,x)$ with $t>0$ and $x \in [0,L]$,
 subject to suitable boundary conditions 
(\eg periodic or Neumann type). The symbol $\xi$ denotes a noise process
which should be thought of as the generalized derivative of some Wiener process
to be specified later on.

Equations of the type (\ref{e:mainequ})  arise in 
the growth of thin films (see \eg \cite{Ra-Ma-Li-Mo-Ha-Sa:00,Si-Pl:94:2nd,Ba-St:95}). 
The function $u(t,\cdot\,)$ describes the graph of a surface at time $t>0$.
Usually these equations are
equipped with a lot of physical parameters, which we set to $1$ for simplicity.
In some models an additional additive nonlinear term $(\partial_x u)^2$ appears.
We can treat that case too, but the analysis is more involved without contributing 
much to the general understanding of the situation, so we do not present it.

It is easy to verify that there exists a value $\nu_c < 0$ such that if 
$\nu < \nu_c$, the equations under consideration present a linear instability. 
We will therefore distinguish in the sequel the {\em stable case} ($\nu > \nu_c$)
from the {\em unstable case} ($\nu \le \nu_c$).
This instability is responsible for the formation of hills, which is frequently 
seen in experiments (see \eg \cite{Ma-Mo-Sa:99} and the references therein).
On the other hand we have  a quadratic nonlinearity that compensates 
this instability. Unfortunately, this nonlinearity makes it difficult to derive uniform bounds on the solution.
Moreover, it is an open problem how to establish bounds in case of a 
two-dimensional square, which is obviously a more  realistic model
than the one-dimensional case we treat in this paper. 
This scenario is similar to the Kuramoto-Sivashinsky equation,
where there are no results for truly two-dimensional domains. 

One very helpful tool in the analysis is the conservation of mass:

\begin{remark}
The quantity $M(u) := \int_0^L u(x)\,dx$ decouples from the rest of the equation.
Therefore, we can assume without loss of generality that
$M\(u(t)\) \equiv 0$ for a solution of (\ref{e:mainequ}). 
The various Sobolev spaces appearing 
in the sequel should be thought of as the orthogonal complement to the
constant function $1$ of the usual spaces.
\end{remark}

The local existence of unique solutions to equations of the type \eref{e:mainequ}
is standard for sufficiently smooth initial conditions. But
the existence of global solutions is much more complicated, and 
was shown in \cite{DB-Gu:02} or \cite{DB-Gu-Ra:02} using a spectral Galerkin method.
Nevertheless the question of uniqueness of global solutions is still open, {\em even in the deterministic case}, as it
is out of reach to show enough regularity.
Therefore the equation does not necessarily generate a Markov semigroup,
and the standard theory for invariant measures (cf. \cite{ZDP})
does not apply.
 
We show in this paper that there exists nevertheless an ``invariant measure'' for \eref{e:mainequ}. To be more precise,  we 
construct a stationary solution $\{u(t),\ t\in\R\}$ such that the distribution 
$\CP^t:=\CL\(u(t)\)\footnote{The symbol $\CL(X)$ will always denote 
the law of the random variable $X$.}$ of $u(t)$ is constant in time. 

Our concept of solutions is a martingale solution of the corresponding mild formulation.
Hence, we allow a change of the underlying  probability space
and consider solutions not of the SPDE, but of the 
corresponding variation of constants formula. 
Since we use spectral Galerkin methods, our approach 
is similar to previous results (see \cite{Fl-Ga:95} or \cite{Ca-Ga:94}) for the stochastic 
Navier-Stokes equation. One of the major differences is
that we are not able to use the theory of Markov semigroups. 
Moreover, we were not able to get any uniform bound (in $t$) on the distribution of
solutions when the driving force $\xi$ is a space-time white noise. We are able to establish such a bound only for stationary solutions.
Therefore, we will construct the stationary process $u$ as a limit of the unique invariant 
stationary solutions to the Galerkin approximations.

The paper is organized as follows. In Section \ref{s:not},
we introduce the spectral Galerkin approximation
and present our main result. The next Section \ref{s:proof}
presents compactness results and the proof of the main result. 
In Sections \ref{s:stable} and \ref{s:unstab}, we will give
a-priori estimates for the solutions. The final Section \ref{s:tech}
contains technical results.

\begin{acknowledge}
D.~B.\ wishes to thank Jean-Pierre Eckmanns mathematical physics group in Geneva for their warm hospitality. M.~H. thanks Jean-Pierre Eckmann, Guillaume van Baalen, and Emmanuel Zabey for several discussions. The latter authors work was partially supported by the Fonds National Suisse. 
\end{acknowledge}

\section{Notation and formulation of the main result}
\label{s:not}

Define the space $\L^2:=\{f\in\L^2([0,L]):\int_0^Lf(x)dx=0\}$
with standard $\L^2$-norm $\|\cdot\|$.
We define $A$ as the linear 
self-adjoint operator in $\L^2$
formally given by
\begin{equ}
  A := -\d_x^4 + \nu \d_x^2\;,
\end{equ}
where the domain of definition $D(A)$  consists 
of all functions  $f\in H^4([0,L])$ satisfying $M(f)=0$ 
and boundary conditions given by the equation.
 We will write $H^4=D(A)$ for short.  
Moreover, it is well-known that $A$ generates 
an analytic semigroup $\{e^{tA}\}_{t\ge0}$, and we 
use the fractional powers of $A$ to define the standard 
fractional Sobolev spaces $H^s$ for $s\ge0$ with dual 
spaces $H^{-s}$.

In the sequel we will need spaces of functions on the whole  real line 
with values in Sobolev spaces. We recall the definition of the  
space $\CC(\R,H^s)$, which is given as the set of all 
functions such that for any $a<b$, the restriction to $[a,b]$ is in
$\CC([a,b],H^s)$. We say that  $f_n$ converges to $f$ in $\CC(\R,H^s)$ 
if and only if $f_n \to f$ in any 
$\CC([a,b],H^s)$ with $a<b$ equipped with the standard maximum norm.
We define the spaces 
$\L^2_\loc(\R,H^s)$ in an analogous way.

We write $\{e_k\}_{k\in\N}$ for a complete 
orthonormal set in $\L^2$ of eigenvectors of $A$ and 
denote by $\Pi_N$ the orthogonal projector onto the 
subspace of $\L^2$ spanned by $e_1,\ldots,e_N$.
Then the $N$th spectral Galerkin approximations 
$u_N$ of \eref{e:mainequ} is given by the solution of
\begin{equ}[e:Galerkin]
  \d_t u_N 
  = A u_N + \Pi_N\d_x^2(\d_x u_N)^2 +\Pi_N\dot{W}\;,
\end{equ}
where $A$ can be interpreted either as an $N\times N$-matrix
or as a differential operator acting on the range of $\Pi_N$.
When considering \eref{e:Galerkin}, 
we will always take initial conditions in the range of $\Pi_N$.
In this equation, $\dot{W}$ is the generalized time-derivative of
a two-sided cylindrical Wiener process $W$ on $\L^2$
with covariance operator $Q$. (See \cite{ZDP1} for the definition of a
cylindrical Wiener process.) We make the following assumption on $Q$:
\begin{assump}
There exist positive numbers $\alpha_k$ and a constant $C$ such that
\begin{equ}
Qe_k=\alpha_k e_k \qquad\text{and}\qquad |\alpha_k| \le C\;,
\end{equ}
for all $k>0$.
\end{assump}
Notice that this assumption covers the case of space-time white noise ($\alpha_k = 1$).
The assumption that $Q$ and $A$ have the same eigenvectors 
implies that we restrict ourselves to translationally invariant noise,
which is also called homogeneous in the physics literature.
This assumption is crucial to verify technical results 
like Lemma \ref{lem:estWA}.

Since \eref{e:Galerkin} is actually a stochastic differential 
equation in $\R^N$ with locally Lipschitz coefficients, 
it is well-known (see \eg \cite{Ha} or \cite{Ar:74}) 
that it  possesses (locally) a strong solution. 
Standard arguments allow to show the following proposition,
the proof of which will be given in \sect{s:stable} below.

\begin{proposition}\label{prop:ex-un}
  For every $N\ge 1$ and any initial condition 
 in $\Pi_N\L^2$, Equation \eref{e:Galerkin} possesses 
 a unique global strong solution. 
  Furthermore, the law of this solution converges in variation norm towards a unique 
  invariant measure $\CQ_N$ which has bounded moments of second order.
\end{proposition}

Consider processes $\{u_N(t)\}_{t\in\R}$ given as stationary solutions of the $N$th 
spectral Galerkin approximation corresponding to the invariant measure $\CQ_N$.  
Hence, $u_N$ satisfies the following
stochastic ODE. 

\begin{equ}[e:Galinv]
  \d_t u_N 
  = A u_N + \Pi_N\d_x^2(\d_x u_N)^2 + \dot W_N\;,
\end{equ}
where $W_N$ is given by 
$W_N(t) = \sum_{k=1}^N \alpha_k e_k\, w_k(t)$ 
with the $\{w_k\}_{k\in\N}$ being a family of independent 
two-sided standard Brownian motions defined on the  probability space underlying $W$.
Since $u_N$ is stationary, we have $\CL\(u_N(t)\)\equiv\CQ_N$ for any $t\in\R$.
By  $\CP_N^{[0,T]}$ we denote the path measure of $\{u_N(t)\}_{t\in[0,T]}$,
and by $\CP_N$ the measure for the whole process $u_N$ in path space.

It is well-known (see \eg \cite{ZDP1}) that, for any pair $t>t_0$, the process $u_N$ satisfies (with probability $1$) the following variation 
of constants formula:
\begin{equation}\label{e:vocuN}
  u_N(t)
  = e^{(t-t_0)A}u_N(t_0)
  +\int_{t_0}^t \partial_x^2 e^{(t-s)A}\Pi_N(\partial_x u_N(s))^2 ds
  +\int_{t_0}^t  e^{(t-s)A} dW_N(s).
\end{equation}
Again, we consider the differential operators either as
operators on the range of $\Pi_N$ or as $N\times N$-matrices.

As our solutions of 
(\ref{e:mainequ})  do not have enough regularity, 
we will focus on mild solutions, which are solutions 
of such integral equations.
Our main result is

\begin{theorem}\label{thm:main}
Consider equation (\ref{e:mainequ}) with periodic or Neumann b.c.
in the stable and only Neumann b.c. in the unstable case. 
Then the family of measures $\{\CQ_N\}_{N\in\N}$ given by \prop{prop:ex-un} is tight on $\L^2$.

Furthermore, for any of its accumulation points $\CQ$, there exists a probability space $(\tilde\Omega,\tilde\CF,\tilde\CP)$, 
a two-sided $Q$-Wiener process $\tilde{W}$, and a {\bf stationary} stochastic
process $\{u(t)\}_{t\in\R}$ with 
$u\in \CC(\R,H^{-3}) \cap \L^2_\loc(\R,H^1)$,
such that $\CL(u(t))\equiv \CQ$ for every $t\in\R$,
and such that
\begin{equ}[e:voc]
  u(t)
  = e^{(t-t_0)A}u(t_0)
  +\int_{t_0}^t \partial_x^2 e^{(t-s)A}(\partial_x u(s))^2 ds
  +\int_{t_0}^t  e^{(t-s)A} d\tilde{W}(s)
\end{equ}
holds for all $t\ge t_0$, $\tilde\CP$-almost surely.
\end{theorem}

We will not focus on optimal regularity, but we could slightly improve the 
regularity of $u$ analogous to Corollary 3.2 and 3.3 of \cite{DB-Gu:02}.
Moreover, we could prove that support of the measure $\CQ$ is concentrated in a
smaller space than  $\L^2$, but we are far from getting enough regularity
to prove pathwise uniqueness.

In the stable case, it is easily possible to prove an analog of Theorem \ref{thm:main} 
with Dirichlet boundary conditions, but we do not enter into details here.

\section{Proof of the main result}
\label{s:proof}

The main step of the proof of \theo{thm:main} is a bound on the logarithmic moments of
$\CQ_N$ that {\it does not} depend on $N$. The main technical difficulty is that It\^o's formula can not be applied to \eref{e:mainequ} since the covariance of our noise is not necessarily trace class. We postpone the proof of \theo{theo:uniformC1} below to sections
\ref{s:stable} and \ref{s:unstab}.

\begin{theorem}\label{theo:uniformC1}
Let $\CQ_N$ be the measure on $\L^2$ invariant for the 
$N$th Galerkin approximation. Then, there exists a constant $C$ 
independent of $N$ such that
\begin{equ}
  \int_{\L^2} \log\(1+\|u\|^2_{\CC^1}\)\,\CQ_N(du) \le C\;,
\end{equ}
uniformly in $N$.
\end{theorem}

Using this result, we turn to the
\begin{proof}[of \theo{thm:main}]
The tightness of the family $\{\CQ_N\}$ follows immediately from \theo{theo:uniformC1} and the compact embedding of $\CC^1$ into $\L^2$. We choose any accumulation point $\CQ$ of $\{\CQ_N\}$ and assume without loss of generality that $\CQ_N$ converges weakly to $\CQ$ in the space of Borel measures on $\L^2$. Denote by $\CP_N$ the law of the (unique in law) stationary process associated to the invariant measure $\CQ_N$ by \prop{prop:ex-un}.

In order to construct the process $u$ appearing in the statement, we first show that the family of measures $\{\CP_N\}$ is tight (it turns out that it is so on the space $\CC(\R,H^{-3}) \cap \L^2_\loc(\R,H^1)$),
and then verify that the limiting process obtained by the usual Prokhorov-Skohorod argument really satisfies the integral equation \eref{e:voc}.

To prove the tightness of the family $\{\CP_N\}$, we
consider $u_N$ as a solution of (\ref{e:vocuN})  with initial 
condition $u_N(0)$ distributed according to $\CQ_N$.
We denote by  $W_A^N(t)$  the stochastic convolution given by
\begin{equ}
W_A^N(t) = \int_0^t e^{A (t-s)}\,dW_N(s)\;,
\end{equ}
and we define $v_N(t) := u_N(t) - W_A^N(t)$. 
The reason is that the stochastic process $v_N$ exhibits 
trajectories with much more time-regularity than $u_N$.
The process $v_N$ is then pathwise a strong solution of the random PDE given by
\begin{equation} \label{e:defvN}
\d_t v_N = A v_N - \d_x^2\Pi_N (\d_x v_N + \d_x W_A^N)^2\;,
\qquad v_N(0)=u_N(0)\;.
\end{equation}

We will need the following technical lemma, 
the proof of which is postponed to sections~\ref{s:stable} and \ref{s:unstab}.
\begin{lemma}\label{lem:apriori}
Fix $\eps,T>0$ and assume that there exists $R>0$ 
such that
\begin{equ}
        \prob\(\|u_N(0)\| > R\) < \eps \qquad\text{for all $N\in\N$.}
\end{equ}
Assume furthermore that $u(0)$ is independent of the Wiener increments for positive times.
Then there exists $\tilde R>0$ independent of $N$ 
such that
\begin{equ}
  \prob\bigl(\|v_N\|_{\CC(0,T,\L^2)} + \|v_N\|_{\L^2(0,T,H^2)} > \tilde R\bigr) 
  < \eps
  \qquad\text{for all $N\in\N$}.
\end{equ}
\end{lemma}

Using this result, we verify the tightness of $\{\CP^{[0,T]}_N\}$ 
 on the space $\CC(0,T,H^{-3}) \cap \L^2(0,T,H^1)$
in a similar way as in \cite[Section 5]{DB-Gu:02}, so we only
briefly sketch the main ideas here.

Given $\eps > 0$, we look for a compact set $K_\eps$ 
such that $\CP^{[0,T]}_N(K_\eps)$ is bounded from below by $1-\eps$ for all $N$.
Combining Theorem \ref{theo:uniformC1} with \lem{lem:apriori}, 
there exists $R$ such that, 
with probability larger than $1-\eps$, $v_N$ lies in 
a ball of radius $R$ of $\CC(0,T,\L^2)\cap\L^2(0,T,H^2)$ and
 $\|v_N(0)\| \le R$. Furthermore,
$W_A^N\to W_A$ in $\CC(0,T,\CC^1)$ $\CP$-a.s. 
Using standard compactness results 
(e.g. \cite[Proposition 1]{Ga:93} or \cite[Proposition 8.4]{ZDP1})
for the integral operator appearing in (\ref{e:vocuN}), 
we can check that the above bounds imply the existence of 
a compact subset $K_\eps^1$ of $C(0,T,H^{-3})$ such that 
$v_N\in K_\eps^1$ with probability larger than $1-\eps$. 
Since $v_N$ is also bounded in $\L^2(0,T,H^2)$
with high probability, we obtain by an interpolation theorem 
(e.g. \cite[Theorem IV.4.1]{Vi-Fu:88})
the existence of a compact set $K_\eps^2\subset\L^2(0,T,H^1)$
such that $v_N\in K_\eps^2$ with probability larger than $1-\eps$.

Hence, $\{\CP^{v_N}\}_{N\in\N}$ is tight on the space 
$\CC(0,T,H^{-3})\cap\L^2(0,T,H^1)$.
By the definition of the projection $\Pi^N$, we 
readily obtain the convergence of $W_A^N=\Pi^N W_A \to W_A$
in $\CC(0,T,H^1)$, as $W_A$ is already in that space.
Combining both arguments, we thus obtain the tightness of the family
$\{\CP_N^{[0,T]}\}_{N\in\N}$ on the space $\CC(0,T,H^{-3})\cap\L^2(0,T,H^1)$.
Since this holds for arbitrary time intervals, 
it is straightforward to extend this to the whole line, 
so $\{\CP_N\}$ is tight on 
$\CC(\R,H^{-3}) \cap \L^2_\loc(\R,H^1)$. We call $\CP^*$ one of its limiting measures
and we obtain a subsequence $\{\CP_{N_k}\}$ that converges weakly to $\CP^*$ in
the abovementioned space.

Now we can use Skohorod's Theorem to obtain a new 
probability space $(\tilde\Omega,\tilde\CF,\tilde\CP)$,
a $Q$-Wiener process $\tilde W$ on that
space, stochastic processes $\tilde u_k$ with laws $\tilde \CP_k=\CP_{N_k}$ solving (\ref{e:vocuN})
with $\Pi_{N_k}\tilde W$ instead of $W_N$,
as well as a stochastic process $\tilde u$ 
with probability distribution $\CL(\tilde u)=\CP^*$ such 
that $\tilde u_k\to\tilde u$ $\tilde\CP$-a.s. in $\L^2_\loc(\R,H^1)\cap\CC(\R,H^{-3})$.
Hence, $\tilde u_k(t)\to \tilde u(t)$ in $H^{-3}$,
and additionally we have $\tilde \CP^t=\CQ$ for all $t\in\R$ by our initial choice of a subsequence.

To show that $\tilde u$ is actually stationary, we first remark that $\tilde{u}_k\to\tilde{u}$
in $\CC(\R,H^{-3})$. Hence, for any choice of $(t_1,\ldots,t_m)\in\R^m$ we readily
obtain in the weak convergence of measures on  $(H^{-3})^m$ that
 $\CL((\tilde{u}_k(t_1),\ldots,\tilde{u}_k(t_m))
 \to \CL((\tilde{u}(t_1),\ldots,\tilde{u}(t_m))$.
Since $\tilde{u}_k$ is stationary, this immediately implies the stationarity of $\tilde u$.

Using the $\tilde\CP$-a.s. convergence
as in \cite[Theorem 3.1]{DB-Gu:02}, 
it is technical but straightforward 
to  verify that $\tilde u$ 
is actually a solution of 
(\ref{e:voc}) with respect to $\tilde W$. 
This completes the proof of Theorem \ref{thm:main}.
\end{proof}

\section{The stable case}
\label{s:stable}

This section provides the postponed proofs of the previous sections in the case of strictly negative $A$. We will discuss the 
necessary changes in order to cover the unstable case in Section~\ref{s:unstab} below.
We start with the

\begin{proof}[of \prop{prop:ex-un}]
The claim follows from \cite{ZDP,Ha}  
if we can show that there exists a constant 
$C$ such that $\expect \|u_N(t)\|^2 \le C$ uniformly in $t$. By It\^o's formula, we have
\begin{equ}[e:estIto]
  d\|u_N\|^2 
  = 2\scal{u_N,Au_N}\,dt 
       + 2\scal{u_N,\d_x^2(\d_x u_N)^2}\,dt 
       + 2\scal{u_N,dW_N(t)} 
       + \sum_{k=1}^N\alpha_k^2 \,dt\;.
\end{equ}
Since $\scal{u_N,\d_x^2(\d_x u_N)^2} = 0$ and $A$ is a strictly negative definite operator, 
the claim follows after integrating $\eref{e:estIto}$ on both sides, 
taking expectations, and applying Gronwall's formula. 
Notice that the bound on the second momenta 
obtained with this procedure diverges with $N$ and it remains an open problem to 
establish a bound independent of $N$ for arbitrary solutions.
\end{proof}

To prove Theorem \ref{theo:uniformC1} for the stable case,
we first verify an $\L^2$-bound.

\begin{theorem}\label{theo:uniform}
Let $\CQ_N$ be the invariant measure on $\L^2$ for the 
$N$th Galerkin approximation. There exists a constant $C$ such that
\begin{equ}
  \int_{\L^2} \log\(1+\|u\|^2\)\,\CQ_N(du) \le C\;,
\end{equ}
for all $N$.
\end{theorem}

\begin{proof}
By (\ref{e:defvN}), the $\L^2$-norm $\|v_N(t)\|^2$  satisfies
\begin{equs}
\d_t \|v_N\|^2 
&=  2\scal{v_N,Av_N} 
     - 2\scal{v_N,\d_x^2 (\d_x v_N + \d_x W_A^N)^2}\\
&=  2\scal{v_N,Av_N} 
     - 2 \int_0^L \d_x(\d_x v_N)^2\,dx 
     - 4 \int_0^L \d_x^2v_N\, \d_x v_N\, \d_xW_A^N\,dx \\
&\qquad    - 2 \int_0^L \d_x^2v_N\, (\d_x W_A^N)^2\,dx \label{e:ap1}\\
&\le -\|\d_x^2 v_N\|^2 
     - \nu \|\d_x v_N\|^2 
     + 4 \|\d_x^2v_N\|\,\|\d_x v_N\| \,\|\d_x W_A^N\|_\infty \\
&\qquad    + 2 \|\d_x^2 v_N\| \,\|\d_x W_A^N\|_4^2\\
&\le - {1\over 4} \|\d_x^2 v_N\|^2 
     - \nu \|\d_x v_N\|^2 + 8\|v_N\|^2\|\d_xW_A^N\|_\infty^4 
     + 4\|\d_xW_A^N\|_4^4 \;.
\end{equs}
Using the Poincar\'e inequality and the fact that we consider only solutions with vanishing mean,
 we see that there exists a positive constant $\alpha$ independent of $N$ (but depen\-ding on $L$) such that
\begin{equ}
\d_t \|v_N\|^2 \le - \alpha \|v_N\|^2 + 8\|v_N\|^2\|\d_xW_A^N\|_\infty^4 + 4\|\d_xW_A^N\|_4^4\;.
\end{equ}
We define now for any interval $[s,t]$ the quantity $W^N_{[s,t]}$ by
\begin{equ}
W^N_{[s,t]} = \int_s^t 8\|\d_xW^N_A(r)\|_\infty^4\,dr\;.
\end{equ}
As a consequence, we have the following {\it a-priori} estimate on the norm of $v_N$:
\begin{equs}
\|v_N(t)\|^2 
&\le e^{-\alpha t + W^N_{[0,t]}}\|v_N(0)\|^2 
     + C\int_0^t e^{-\alpha (t-s) + W^N_{[s,t]}}\(1+\|\d_xW_A^N(s)\|_4^4\)\,ds\\
&\le e^{-\alpha t + W^N_{[0,t]}}\|v_N(0)\|^2 + C e^{W^N_{[0,t]}} (W^N_{[0,t]}+t)\;.\label{e:ap2}
\end{equs}
Since $u_N = v_N + W_A^N$ and $W_A^N(0)=0$ we obtain for 
some $\eps>0$ fixed later on:
\begin{equs}
\|u_N(t)\|^2 
&\le (1+\eps)\|v_N(t)\|^2+C\|W_A^N(t)\|^2\\
& \le (1+\eps)e^{-{\alpha} t + W^N_{[0,t]}}\|u_N(0)\|^2 
       + C e^{W^N_{[0,t]}} (W^N_{[0,t]}+t) 
       + C \|W^N_A(t)\|^2.
\end{equs}
Note that the constants may depend on $\eps$.

The problem at this point is that the exponential moment of 
the random variable $W^N_{[0,t]}$ is infinite.
We therefore take logarithms on both sides, yielding
\begin{equa}\label{e:s.o.}
\log\(1 + \|u_N(t)\|^2\) 
&\le \log\Big((1+\eps)\, e^{-\alpha t + W^N_{[0,t]}}\(1+\|u_N(0)\|^2\)  \\
&\qquad\qquad+ C e^{W^N_{[0,t]}} \(W^N_{[0,t]}+t\) 
+ C \|W^N_A(t)\|^2+1\Big)\;.
\end{equa}

Using \lem{lem:estWA} it is now easy to verify that 
we can apply \lem{lem:estimate} (with a constant $K$ independent of $N$)
to the right-hand side of (\ref{e:s.o.}), where we take the conditional expectation 
w.r.t.\ $u_N(0)$. Hence,
\begin{equs} \label{e:estlogu}
\expect \Big(\big(\log&\(1 + \|u_N(t)\|^2\)\big)^2\,\Big|\,u_N(0)\Big) 
\le \Bigl(\log\(1 + \|u_N(0)\|^2\)\Bigr)^2 \\
& +2\(\eps+\log(1+\eps)- \alpha t+\expect W^N_{[0,t]}\)\cdot \log\(1 + \|u^N(0)\|^2\) + C\;, 
\end{equs}
for some constant $C$ depending on $\eps$ and on the parameters of the problem, but not on $N$.

At this point, we choose first  $t$ sufficiently small such  that 
\begin{equ}
\expect W^N_{[0,t]} \le {\alpha t \over 2}\;.
\end{equ}
This can be done uniformly in $N$ by \lem{lem:estWA}.
Then fix $\eps$ so small such that 
$$\eps+\log(1+\eps)\le {\alpha t\over 4}.
$$
Taking expectations on both sides of \eref{e:estlogu} and using the stationarity of $u_N(t)$, we have
\begin{equ}
  \expect \log\(1 + \|u_N(0)\|^2\) \le {C \over \alpha t}\;
\end{equ}
for fixed $t$ sufficiently small,
therefore concluding the proof of \theo{theo:uniform}.
\end{proof}

\begin{remark}
\theo{theo:uniform} establishes only  a uniform bound (in $t$) 
for stationary solutions of our problem. 
Estimates like (\ref{e:s.o.}) or (\ref{e:estlogu}) are not sufficient 
to get uniform bounds for arbitrary solutions, and this question 
remains open.
\end{remark}

Let us now turn to the

\begin{proof}[of \theo{theo:uniformC1}]
Using  \eref{e:ap1} we obtain after integration
\begin{equs}
\int_0^1\|\d_x^2 v_N(t)\|^2 dt
&\le C\int_0^1\|v_N(t)\|^2(1+\|\d_x W_A^N(t)\|^4_\infty)\,dt
\label{e:estH2}\\
&\quad    +C\int_0^1\|\d_x W_A^N(t)\|^4_4 \,dt + C\|v_N(0)\|^2\;.
\end{equs}
Using Young's inequality and the Sobolev embedding of $H^1$ into $\L^\infty$,
we have the bound
$$\|u_N\|_{\CC^1}^2
  \le C\|\d_x^2 v_N\|^2+C\|\d_x W_A^N\|^4_\infty+C\;.
$$
This yields
\begin{equs}
\int_0^1&\|u_N(t)\|_{\CC^1}^2 \, dt
 \le C\int_0^1\| v_N(t)\|^2\(1+\|\d_x W_A^N(t)\|^4_\infty\)\,dt\\
&\quad    +C\int_0^1\|\d_x W_A^N(t)\|^4_4\, dt+C + C\|u_N(0)\|^2\\
&\le  C\int_0^1 \left(e^{-\alpha t + W^N_{[0,t]}}\|u_N(0)\|^2 +  e^{W^N_{[0,t]}} \(W^N_{[0,t]}+t\)\right)\\
&\quad\times\(1+\|\d_x W_A^N(t)\|^4_\infty\)\,dt
 +C\int_0^1\|\d_x W_A^N(t)\|^4_4\,dt+C + C\|u_N(0)\|^2\\
&\le  C \left(\bigl(e^{ W^N_{[0,1]}}+1\bigr)\|u_N(0)\|^2 +  e^{W^N_{[0,1]}} \(W^N_{[0,1]}+1\)+1\right)\(1+W^N_{[0,1]}\)\;,
\end{equs}
where we used (\ref{e:ap2}).
Using Theorem \ref{theo:uniform} and \lem{lem:estWA}, we immediately obtain
$$
\expect\log\left(\int_0^1\|u_N(t)\|_{\CC^1}^2 dt+1\right)
\le C.
$$
Finally, Jensen's inequality and the stationarity of $u_N$ 
yield
\begin{equs}
\expect\log\left(\|u_N(0)\|_{\CC^1}^2 +1\right)
& = \expect\int_0^1\log\left(\|u_N(t)\|_{\CC^1}^2+1\right)\,dt\\
&\le \expect\log\left(\int_0^1\|u_N(t)\|_{\CC^1}^2 dt+1\right)\le C\;,
\end{equs}
concluding the proof of \theo{theo:uniformC1}.
\end{proof}

We now turn to the proof of \lem{lem:apriori}. 
This proof will not use the strict negativity of $A$ and is thus still valid in the unstable case. Since we need this bound only for a fixed time interval $[0,T]$, 
we can bound the terms in a rather crude way.
\begin{proof}[of \lem{lem:apriori}]
Define 
$$ W_T^N:=\sup_{t\in[0,T]}\|\d_x W_A^N(t)\|_\infty^4\;.
$$
Using the factorization method and Sobolev embedding 
it is straightforward to check that $\expect W_T^N<C$
uniformly in $N$.
This result is established completely analogous to
\cite[Lemma 5.1]{Bl-MP-Sc:01}.
Note that the uniformity in $N$ is not trivial, as
the family $\{\Pi^N\}_{N\in\N}$ is not uniformly bounded 
as operators on $L^\infty$ or $\CC^0$.

Using this and the assumption on $\|u_N(0)\|$, we see that
for every $\eps>0$ there is an $R>0$ such that
\begin{equ}[e:boundW_T]
 \prob\(W_T^N<R\quad\text{and}\quad \|u_N(0)\|<R\)>1-\eps\;.
\end{equ} 
Combining \eref{e:boundW_T} and \eref{e:ap2}, we see that with probability larger than $1-\eps$ one has
\begin{equs}
\|v_N(t)\|^2
&\le
  e^{W_T^N}\|v_N(0)\|^2+Ce^{W_T^N}(W_T^N+T)\\
&\le e^R R^2+Ce^R(R+T)=:R' \;,
\end{equs} 
for any $t\in[0,T]$. Using \eref{e:estH2} in the same way, we get
\begin{equs}
\int_0^T\|\d_x^2 v_N(t)\|^2\;dt
&\le
C \|v_N(0)\|^2+C\int_0^T \|v_N(t)\|^2(W_T^N+1)dt+CTW_T^N\\
&\le C(R+TR'(R+1)+TR)\;,
\end{equs} 
with probability larger than $1-\eps$, thus concluding the proof of \lem{lem:apriori}.
\end{proof}

\section{The unstable case}
\label{s:unstab}

This section deals with the case where the operator 
$A$ is no longer strictly negative definite. In order to treat this case,
we make use of a trick that was used in \cite{NST85,Co-Ec-Ep-St:93}
to get bounds on the deterministic Kuramoto-Sivashinsky equation.
It turns out that the present model is sufficiently close to that equation to make that trick go through.
Nevertheless, we can only treat Neumann boundary conditions
(which is the same as considering the restriction on $[0,L]$ of functions 
that are even and periodic with period $2L$). 
In a similar way we can treat also Dirichlet boundary conditions,
but periodic boundary conditions are still open.
 
Most of the proofs are similar to the previous section, so we will only state the main differences.
Instead of defining $v_N$ as previously, we define $v_N$ by
\begin{equ}[e:defv]
 v_N(t) = u_N(t) - W_A^N(t) - \Phi_N\;,
\end{equ}
where $\Phi_N= \Pi_N \Phi$ for some function $\Phi$ 
 to be chosen later and $W_A^N$ is the stochastic convolution defined 
in the previous section. The stochastic process $v_N$ then satisfies the following random PDE:
\begin{equ}[e:vunstable]
 \d_t v_N = A v_N + A\Phi_N - \Pi_N \d_x^2 (\d_x v_N + \d_x W_A^N + \d_x \Pi_N\Phi)^2\;.
\end{equ}
We can rewrite this as
\begin{equ}
\d_t v_N = \tilde A v_N 
   - \d_x^2 (\d_x v_N + \d_x W_A^N)^2 
   - \d_x^2 (\d_x \Pi_N\Phi)^2 + A\Phi 
   - 2\d_x^2(\d_x \Phi\,\d_x W_A)\;,
\end{equ}
where the operator $\tilde A$ is defined as
\begin{equ}
 \tilde A v = A v - 2 \d_x^2 (\d_x\Phi\,\d_x v)\;.
\end{equ}
Using exactly the same technique as in the previous section,
we see that in order to get uniform bounds on the Galerkin approximations of $v$,
it suffices to find a smooth function $\Phi$ such that
\begin{equ}[e:Atilde]
 \scal{v,\tilde Av} \le -c\|\d_x^2 v\|^2
\end{equ}
for some constant $c>0$. 
Using this function, it is easy to verify that the assertions 
of Proposition \ref{prop:ex-un} and 
Theorems \ref{theo:uniformC1} and \ref{theo:uniform} hold in the unstable case, too.
The only major changes appear in the values of the constants, 
which do now depend on the choice of $\Phi$.  
We will therefore not go through the proofs of these assertions for the unstable case, but
we will sketch how to find a function $\Phi$ such that $\tilde A$ is strictly negative definite.

Integrating by parts, we see that the bilinear form \eref{e:Atilde} can be written as
\begin{equ}[e:bilin]
 \scal{v,\tilde Av} = - \|\d_x^2 v\|^2 - \nu \|\d_x v\|^2 - \scal{\d_x v, (\d_x^2\Phi)\,\d_x v}\;,
\end{equ}
where $\nu$ is negative. The problem is therefore reduced to finding a smooth periodic function $\Phi$ such that the Schr\"odinger operator 
\begin{equ}
 \CH_\Phi = -\textstyle{1\over 2}\d_x^2 + \d_x^2 \Phi(x)\;,
\end{equ}
with {\it Dirichlet} boundary conditions satisfies $\scal{u,\CH_\Phi u} \ge |\nu|\|u\|^2$ for all functions $u$ in its domain. 
The idea appearing in \cite{NST85} is to choose $\Phi$ such that, away from the boundary, $\d_x^2\Phi$ is for all practical purposes constant and sufficiently large (say equal to about $2|\nu|$). The problem is that, in order for \eref{e:vunstable} to hold, $\Phi$ has to belong to $D(A)$ and must therefore satisfy the same boundary conditions as $u$. As a consequence $\d_x^2\Phi$ must satisfy $\int_0^L \d_x^2\Phi(s)\,dx = 0$, which is of course impossible for a constant (non-zero) function. Looking at \eref{e:bilin}, we notice that $\d_x^2\Phi(x) = 2|\nu|\(1-\delta(x)\)$ would formally fit our needs, since the delta-peak is integrated against $\d_x v$, which vanishes at the boundaries, due to the Neumann conditions. The function $\Phi$ obtained this way does of course not belong to $D(A)$, so we look for an approximation of it which is more regular.

Since $\Phi$ satisfies Neumann boundary conditions, it is natural to write it as
\begin{equ}[e:reprPhi]
 \Phi(x) = 2|\nu|\sum_{n=1}^\infty \phi_n\,\cos\Bigl({2\pi n\over L}\Bigr)\;.
\end{equ}
(The sum starts at $1$ because we are interested only in functions with vanishing mean.)
If we choose $\phi_n = 2n^{-2}$, we see that $\d_x^2\Phi$ is given by $\d_x^2\Phi(x) = 2|\nu|\(1-\delta(x)\)$, which is what we would like to approximate. In order to get a regular function, we define
\begin{equ}[e:defPhi]
\psi_n := n^2 \phi_n = \cases{2&for $n\le 2n_*$,\cr 0 &for $n>2n_*$,}
\end{equ}
where $n_*$ is some (sufficiently large) constant to be fixed later on. 
With this definition, we have:

\begin{proposition}\label{prop:Phi}
For every $L,C>0$, there exists a value $n_* > 0$ such that the quadra\-tic form $\CH_\Phi$ with $\Phi$ defined as in \eref{e:reprPhi} and \eref{e:defPhi}, satisfies
\begin{equ}
 \int_0^L u(x) \(\CH_\Phi u\)(x)\,dx \ge C\|u\|^2\;,
\end{equ}
for every $u$ in the domain of $\CH_\Phi$.
\end{proposition}

\begin{remark}
Notice that the function $\Phi$ defined by \eref{e:reprPhi} and \eref{e:defPhi} is actually analytic, so the expressions appearing in \eref{e:vunstable} and containing $\Phi$ can all be bounded uniformly in $N$ (not in $n_*$ of course, but $n_*$ is chosen independently of $N$).
\end{remark}

\begin{remark}
As in \cite{Co-Ec-Ep-St:93} we could choose a slow decay of $\psi_n$ for $n>2n_*$ to 
optimize the $L$-dependence of our bound, but we neglected this for simplicity. 
\end{remark}

\begin{proof}
Applying the arguments of \cite[Prop.~2.1]{Co-Ec-Ep-St:93}, we see that it suffices to show that the quantity
\begin{equ}
\Gamma := \sum_{k>m>0}{|\psi_{k+m}-\psi_{k-m}|{\rlap{${}^2$}}\over E_k E_m}\;,\quad\text{with}\quad E_n = \alpha n^2 \;,\quad \alpha= {2\pi^2 \over L^2}\;,
\end{equ}
can be made arbitrarily small by choosing $n_*$ sufficiently large. The only non-vanishing terms of this sum are those where $0 \le k-m \le 2n_*$ and $k+m \ge 2n_*$. We can estimate these terms by
\begin{equs}
        \Gamma &\le {4 \over \alpha^2} \sum_{m=1}^{n_*-1} \sum_{k=2n_*-m}^{2n_*+m} {1\over m^2 k^2} + {4 \over \alpha^2} \sum_{m=n_*}^{\infty} \sum_{k=m}^{2n_*+m} {1\over m^2 k^2} \\
&\le {8 \over \alpha^2 n_* }  \sum_{m=1}^{n_*-1} {1 \over m^2} + {8 \over \alpha^2 n_* }  \sum_{m=n_*}^{\infty} {1 \over m^2} \le {4 \pi^2 \over 3\alpha^2 n_* }\;.
\end{equs}
In both sums, we used the fact that $k$ is larger than $n_*$ and that there are less than $2n_*$ terms in the inner sum.
Thus, $\Gamma$ can clearly be made arbitrarily small by choosing $n_*$ sufficiently large. This proves \prop{prop:Phi} and concludes our exposition of the unstable case.
\end{proof}

\section{Technical estimates}
\label{s:tech}

In this section, we prove the two technical estimates required for the proof of \theo{theo:uniform} above.
\begin{lemma}\label{lem:estWA}
There exists a constant $C$ independent of $N$ such that
\begin{equ}
\expect \|\d_x W_A^N(t)\|_\infty^4 \le C t^{1/8}
\end{equ}
for all $t \le 1$.
\end{lemma}

\begin{remark}
The power $1/8$ in the above lemma is not optimal but it is sufficient for our needs.
All we need is $\expect \|\d_x W_A^N(t)\|_\infty^4=o(1)$ uniformly in $N$.
\end{remark}

\begin{remark}
The constant in the above lemma depends only on the
coefficients of the problem and the  bound on the $\alpha_k$.
It is possible to allow for slowly growing $\alpha_k$, using the Sobolev embedding
of $\L^\infty$ into the fractional Sobolev space $W^{s,p}$ for $sp>1$. 
\end{remark}

\begin{proof}
For $f \in \L^\infty([0,L])$ with vanishing mean, denote by $\{f_k\}_{k\in\N}$
its Fourier coefficients. Since the eigenfunctions $e_k$ of $A$
are uniformly bounded in $\L^\infty$,
we have the following estimate on $\|f\|_\infty$:
\begin{equ}
\|f\|_\infty \le C \sum_{k=1}^\infty {|f_k|} 
\le C \Bigl(\sum_{k=1}^\infty |k|^{1/2}|f_k|^{4/3}\Bigr)^{3/4}\Bigl(\sum_{k=1}^\infty |k|^{-3/2}\Bigr)^{1/4}
\le C \|Kf\|_4\;,
\end{equ}
where $K$ is the operator that acts on Fourier coefficients as $(Kf)_k = |k|^{3/8}f_k$. Here we used the usual isometry between $\L^p$ and $\ell^q$ for $p^{-1} + q^{-1} = 1$.

Denote by $\lambda_k$ the eigenvalues corresponding
to the eigenfunctions $e_k$. By the definition of $A$, 
there exist constants $c_i$ such that
\begin{equ}
c_1 k^4 \le -\lambda_k \le c_2 k^4\;,\qquad |\(K\d_x e_k\)(x)|^2 \le c_3 k^{11/4}\;.
\end{equ}
With these notations, $\(K\d_x W_A^N\)(t,x)$ (with fixed values of $x$ and $t$) are centered Gaussian
random variables given by
\begin{equ}
\(K\d_x W_A^N\)(t,x) = \sum_{k=1}^N \alpha_k \(K \d_x e_k\)(x)\, \int_0^t e^{-\lambda_k(t-s)}\,dw_k(s)\;,
\end{equ}
with independent Wiener processes $w_k(t)$. The variance of $\(K\d_x W_A^N\)(t,x)$ is thus bounded by
\begin{equs}
\expect \bigl|\bigl(K\d_x W_A^N\bigr)(t,x)\bigr|^2 
&= \sum_{k=1}^N \alpha_k^2 |\(K\d_x e_k\)(x)|^2 \int_0^t e^{2 \lambda_k(t-s)}\,ds\\
&\le C\cdot\sum_{k=1}^\infty  |k|^{11/4}  \int_0^t e^{-c_1 k^4 s}ds 
 \le \sum_{k=1}^\infty {C \over k^{5/4}} \(1-e^{-c_2 k^4 t}\)\;.
\end{equs}
We now take some $k_*$ to fixed later and split the sum into two parts:
\begin{equ}
\expect |\(K\d_x W_A\)(x,t)|^2 \le C\sum_{k=1}^{k_*-1} k^{11/4} t + C\sum_{k=k_*}^\infty {1 \over k^{5/4}} \le C k_*^{15/4} t + {C \over k_*^{1/4}}\;.
\end{equ}
Choosing $k_* \approx t^{-1/4}$, we have the estimate $\expect |\(K\d_x W_A^N\)(x,t)|^2 \le C t^{1/16}$. 
Since the random variables $\(K\d_x W_A^N\)(x,t)$ are Gaussian for all fixed values $(x,t)$, we have 
\begin{equ}
 \expect \|\d_x W_A^N\|_\infty^4 \le C\expect \|K\d_x W_A^N\|_4^4
= \int_0^L \expect \(K\d_x W_A\)(x,t)^4\,dx \le C t^{1/8}\;.
\end{equ}
\end{proof}

\begin{lemma}\label{lem:estimate}
For every $\eps > 0$ and $K>0$ there exists a constant  $C$ depending only on $\eps$ and $K$, 
such that for any two random variables $W_1$ and $W_2$ with 
$\expect\(|W_1|^2 + |W_2|^2\) \le K$ we obtain
\begin{equ}
\expect \Bigl(\log\(x e^{W_1} + e^{W_2}\)\Bigr)^2 \le \(\log(x)\)^2 + 2(\eps + \expect W_1)\log x + C\;,
\end{equ}
for every $x \ge 1$.
\end{lemma}

\begin{proof}
Expanding the square, we get
\begin{equs}
\expect \Bigl(\log\(x e^{W_1} + e^{W_2}\)\Bigr)^2 &= (\log x)^2 + 2\log x\, \Bigl(\expect W_1 + \expect \log\(1 + e^{W_2 - W_1}/x\)\Bigr) \\
&\quad + \expect\Bigl(W_1 + \log\(1 + e^{W_2 - W_1}/x\)\Bigr)^2\;.
\end{equs}
Since we assumed $x\ge 1$, we have
\begin{equs}
  \expect\Bigl(W_1 + \log\(1 + e^{W_2 - W_1}/x\)\Bigr)^2 
  &\le 2\(\expect W_1^2 + \expect (W_2 - W_1)^2 + 1\)\\
  &\le 6K+2\;.
\end{equs}
It thus suffices to show that there exists $x_0 >0$ 
depending only on $K$ and on $\eps$ such that
\begin{equation}\label{e:Ex}
  E_x := \expect\log\Bigl(1 + {e^{W_2 - W_1}\over x}\Bigr) \le \eps
\end{equation}
for $x$ larger than $x_0$. 

To verify (\ref{e:Ex}) consider arbitrary $\eps>0$.  
We define $y = |W_2 - W_1|$ and denote the probability distribution 
on $\R_+$ of $y$ by  $\prob$.
Now choose $y_0 > 1$ large enough such that
\begin{equ}
 \int_{y_0}^\infty y\,\prob(dy) \le \eps\;,\qquad \int_{y_0}^\infty \prob(dy) \le \eps\;.
\end{equ}
We choose $y_0 = 1+(\expect y^2)/\eps$. Now define
\begin{equ}
 x_0 = {e^{1+2K/\eps}\over \eps}\ge {e^{y_0}\over \eps} \;.
\end{equ}
We thus have
\begin{equs}
 E_x &= \int_0^{y_0} \log\Bigl(1 + {e^y \over x}\Bigr)\,\prob(dy) + \int_{y_0}^{\infty} \log\Bigl(1 + {e^y \over x}\Bigr)\,\prob(dy) \\
&\le \log\Bigl(1+{e^{y_0} \over x}\Bigr) + \int_{y_0}^\infty \log\(1 + \eps e^{y - y_0}\)\,\prob(dy) \\
&\le \eps +  \int_{y_0}^\infty \log\((1+\eps)e^{y - y_0}\)\,\prob(dy)\\
&\le \eps + \eps\log(1+\eps) + \int_{y_0}^\infty (y - y_0)\,\prob(dy)
\le 2\eps + \eps\log(1+\eps)\;.
\end{equs}
The claim follows immediately.
\end{proof}


\bibliographystyle{myalph}

\markboth{\sc \refname}{\sc \refname}
\bibliography{refs}
\end{document}